# A new class of exact solution of two-dimensional incompressible vortex motion

Zheng Ran

(Shanghai Institute of Applied Mathematics and Mechanics, Shanghai University, Shanghai, 200072)

**Abstract:** At present in the fluid mechanics, mostly one like to use the vortex as a basic physical quantity, such that some exact solutions is based on the vorticity evolution equation. For the vortex flow problem with axisymmetry, it is well known that there exists in the circulation as a mechanical system of basic physical quantities. In this paper, from the basic dynamic equation of the mechanical system, the self-similar solution, eigenvalue system into a self- consistent, new exact solutions of the two-dimensional circular symmetric vortex flow are obtained, and some further comparison are made with the known exact solutions.

## 0. Introduction

Exactly solvable models are one of very important research field in physics. It is well known that in the quantum mechanical [1,2,3], statistical mechanics[4] and fluid mechanics[5,6], we have already found many kinds of exactly solvable systems. It is useful, both from the mathematical methods, and discussing the nature of the physical system. In this paper, using the generalized method of self-similarity solution, we introduce an eigenvalue system, to study the two-dimensional is axisymmetric incompressible vortex flow, and found some new solutions. Some comparisons are made with the Neufville [7] solution.

## 1. The dynamical equations

It often convenient to use a cylindrical coordinate system $(r, \theta, z)$ to describing columnar vortices and vortex ring with the velocity $\vec{V} = (u, v, w)$, and vorticity $\vec{\omega} = (\omega_r, \omega_\theta, \omega_z)$. The vorticity components are given by [5,6]

$$\omega_r = \frac{1}{r}\frac{\partial w}{\partial \theta} - \frac{\partial v}{\partial z}, \tag{1}$$

$$\omega_\theta = \frac{\partial u}{\partial z} - \frac{\partial w}{\partial r}, \tag{2}$$

$$\omega_z = \frac{1}{r}\frac{\partial (rv)}{\partial r} - \frac{1}{r}\frac{\partial u}{\partial \theta}. \tag{3}$$

The continuity equation reads:

$$\frac{1}{r}\frac{\partial (rv)}{\partial r} + \frac{1}{r}\frac{\partial v}{\partial \theta} + \frac{\partial w}{\partial z} = 0. \tag{4}$$

It well known that the velocity and vorticity, and hence their governing equations can be expressed in terms of two scalar functions, one is the Stokes stream function ψ, the other is

$$\Gamma = rv. \tag{5}$$

Differing from the circulation around a circle centered at $r = 0$ by a factor $1/2\pi$. Consequently, the definition of vorticity yield

$$\omega_r = -\frac{1}{r}\frac{\partial \Gamma}{\partial z}, \tag{6}$$

$$\omega_\theta = \frac{1}{r}\frac{\partial \Gamma}{\partial r}, \tag{7}$$

$$\omega_z = -\frac{1}{r}\frac{\partial}{\partial r}\left(\frac{1}{r}\frac{\partial \psi}{\partial r}\right) - \frac{1}{r}\frac{\partial \psi}{\partial z}. \tag{8}$$

The vorticity transport equations can be cast to

$$\frac{D\Gamma}{Dt} = \nu\left[r\frac{\partial}{\partial r}\left(\frac{1}{r}\frac{\partial \Gamma}{\partial r}\right) + \frac{\partial^2 \Gamma}{\partial z^2}\right], \tag{10}$$

$$\frac{D}{Dt}\left(\frac{\omega_\theta}{r}\right) = \nu\left(\nabla^2 + \frac{2}{r}\frac{\partial}{\partial r}\right)\left(\frac{\omega_\theta}{r}\right) + \frac{1}{r^4}\frac{\partial \Gamma^2}{\partial z}, \tag{11}$$

where

$$\frac{D}{Dt} = \frac{\partial}{\partial t} + u\frac{\partial}{\partial r} + v\frac{\partial}{\partial \theta} + w\frac{\partial}{\partial z}, \tag{12}$$

$$\nabla^2 = \frac{1}{r}\frac{\partial}{\partial r}\left(r\frac{\partial}{\partial r}\right) + \frac{1}{r^2}\frac{\partial^2}{\partial \theta^2} + \frac{\partial^2}{\partial z^2}. \tag{13}$$

These equations govern the azimuthal and meridional motions, respectively.

It is therefore appropriate here to examine the simplest case of the exact solution. An inspection of these equations with $[1]\frac{\partial}{\partial \theta} = 0$, $[2]v_r = 0$, $[3]v_z = 0$, and also

$$v_\theta = v_\theta(r,t), \tag{14}$$
$$p = p(r,t). \tag{15}$$

Consequently, the above equations yield:

$$\frac{\partial \Gamma}{\partial t} = \nu\left(\frac{\partial^2 \Gamma}{\partial r^2} - \frac{1}{r}\frac{\partial \Gamma}{\partial r}\right). \tag{16}$$

This is the classical solution of an isolated line vortex in an otherwise undisturbed fluid was first discussed by Oseen (1911) for unsteady flow. A complete set of similarity solutions has been given by Neufville [7].

## 3.Similarity analysis

This equation admits a self-similar solution with

$$\eta = \frac{r}{l(t)}. \tag{17}$$

where $l = l(t)$ is a physical quantity must be determined.

We now may write

$$\Gamma(r,t) = A(t)f(\eta). \tag{18}$$

Introducing the similarity variable (17) and employing the relationship (8) gives

$$\frac{dA}{dt}f - \frac{A}{l}\frac{dl}{dt}\eta\frac{df}{d\eta} = \frac{\nu A}{l^2}\left(\frac{d^2f}{d\eta^2} - \frac{1}{\eta}\frac{df}{d\eta}\right). \tag{19}$$

It is convenient to divide by $\frac{\nu A}{l^2}$, so that the transformed equation reduces to

$$\frac{d^2f}{d\eta^2} + \left(\frac{l}{v}\frac{dl}{dt}\eta - \frac{1}{\eta}\right)\frac{df}{d\eta} - \frac{l^2}{vA}\frac{dA}{dt}f = 0. \tag{20}$$

Now, since the coefficient of the term is time independent, self-preserving solutions of the type sought here are possible only if the terms in above equation satisfying

$$\frac{l}{v}\frac{dl}{dt} = a, \tag{21}$$

$$-\frac{l^2}{vA}\frac{dA}{dt} = b. \tag{22}$$

The argument $(a, b)$ denotes two constants. Substituting into the equation (20) leads immediately to the transformed equation

$$\frac{d^2f}{d\eta^2} + \left(a\eta - \frac{1}{\eta}\right)\frac{df}{d\eta} + bf = 0. \tag{23}$$

The substitution

$$f(\eta) = \eta^{\frac{1}{2}}\exp\left(-\frac{a}{4}\eta^2\right)\phi(\eta). \tag{24}$$

brings equation (23) to the canonical ( or normal )form : (Appendix 1)

$$\frac{d^2\phi}{d\eta^2} + \left\{b - \frac{1}{4}a^2\eta^2 - \frac{3}{4\eta^2}\right\}\phi = 0. \tag{25}$$

This equation, together with the boundary conditions, clearly constitutes a characteristic value problem.

Letting

$$\eta = \delta x. \tag{26}$$

where $\delta$ is a constant must be determined, we can rewrite the differential equation (25) in the form

$$\frac{d^2\phi}{dx^2} + \left\{b\delta^2 - \frac{1}{4}a^2\delta^4 x^2 - \frac{3}{4x^2}\right\}\phi = 0. \tag{27}$$

From the details of the Appendix 2, it follows that

$$b\delta^2 = 4n + 2\alpha + 2, \tag{28}$$

$$\frac{1}{4}a^2\delta^4 = 1, \tag{29}$$

$$-\frac{3}{4} = \frac{1-4\alpha^2}{4}. \tag{30}$$

Where n = 0,1,2,3, …… is an integer.
The corresponding solutions for parameters are

$$\alpha = 1, \tag{31}$$

$$\delta = \sqrt{\frac{2}{a}}, \tag{32}$$

$$b\delta^2 = 4(n + 1). \tag{33}$$

For a given n, we find that

$$\phi(x) = x^{\alpha+\frac{1}{2}}\exp\left(-\frac{1}{2}x^2\right)L_n^{(\alpha)}(x^2). \tag{34}$$

where $L_n^{(\alpha)}(x)$ denote the generalized Laguerre polynomial.
From the equations (11) and (12) it follows immediately that

$$l^2 = 2av(t + t_0), \tag{35}$$

$$A = A_0(t + t_0)^{-(n+1)}. \tag{36}$$

where $t_0, A_0$ are two constants.

We may express the solution of the equation (16) in the form:

$$\Gamma_n(r,t) = A_0 \left(\frac{a}{2}\right)^{\frac{3}{4}} (t+t_0)^{-(n+1)} \eta^2 \exp\left(-\frac{a}{2}\eta^2\right) L_n^{(1)}\left(\frac{a}{2}\eta^2\right). \tag{37}$$

It is convenient to introduce a new variable as

$$\zeta = \frac{a}{2}\eta^2. \tag{38}$$

Substituting for $\zeta$ from the equation (37), we obtain

$$\Gamma_n(r,t) = A_0 \left(\frac{2}{a}\right)^{\frac{1}{4}} (t+t_0)^{-(n+1)} \zeta \exp(-\zeta) L_n^{(1)}(\zeta). \tag{39}$$

It follows the definition of $\Gamma$, we have

$$v_\theta = \frac{\Gamma}{2\pi r}. \tag{40}$$

The solution for velocity is

$$v_n(r,t) = \frac{A_0}{2\pi\sqrt{2a\nu}} \left(\frac{a}{2}\right)^{\frac{1}{4}} (t+t_0)^{-\left(n+\frac{3}{2}\right)} \zeta^{\frac{1}{2}} \exp(-\zeta) L_n^{(1)}(\zeta). \tag{41}$$

## 4. The solution of vorticity

The vorticity component is given by

$$\omega = \omega_z = \frac{1}{r}\frac{\partial(rv_\theta)}{\partial r}. \tag{42}$$

Based on the definition of $\Gamma$, we have

$$2\pi\omega = \frac{1}{r}\frac{\partial \Gamma}{\partial r}. \tag{43}$$

Making further use the recurrence relations satisfied by the generalized Laguerre function, we find that

$$\omega_n = \frac{A_0}{4\pi\nu} \left(\frac{2}{a}\right)^{\frac{1}{4}} (t+t_0)^{-(n+2)} (n+1) \exp(-\zeta) L_{n+1}^{(0)}(\zeta). \tag{44}$$

or

$$\omega_n = \frac{A_0}{4\pi\nu} \left(\frac{2}{a}\right)^{\frac{1}{4}} (t+t_0)^{-(n+2)} \exp(-\zeta) \left[(n+1) L_n^{(0)}(\zeta) - \zeta L_n^{(1)}(\zeta)\right]. \tag{45}$$

## 5. Comparison with presvious Neufville's solution

In our new solution, the mode $n = 0$ is the Taylor vortex:

$$\omega_0 = \frac{A_0}{4\pi\nu} \left(\frac{2}{a}\right)^{\frac{1}{4}} (t+t_0)^{-2} \exp(-\zeta) L_1^{(0)}(\zeta). \tag{46}$$

and

$$\Gamma_0(r,t) = A_0 \left(\frac{2}{a}\right)^{\frac{1}{4}} (t+t_0)^{-1} \zeta \exp(-\zeta). \tag{47}$$

$$v_0(r,t) = \frac{A_0}{2\pi\sqrt{2a\nu}} \left(\frac{a}{2}\right)^{\frac{1}{4}} (t+t_0)^{-\left(\frac{3}{2}\right)} \zeta^{\frac{1}{2}} \exp(-\zeta). \tag{48}$$

As we know that Neufville [7] also gave a complete set of similarity solutions as follows:

$$\bar{\omega}_n = d_0^n (t + t_0)^{-(n+1)} exp(-\zeta) L_n^{(0)}(\zeta). \quad (49)$$

$$\bar{v}_n(r,t) = \sqrt{\frac{v}{(t+t_0)}} d_0^n (t + t_0)^{-n} \zeta^{\frac{1}{2}} exp(-\zeta) L_{n-1}^{(1)}(\zeta). \quad (50)$$

$$\bar{\Gamma}_n(r,t) = 4\pi v d_0^n (t + t_0)^{-n} \zeta exp(-\zeta) L_{n-1}^{(1)}(\zeta). \quad (51)$$

Where $d_0$ is the constant introduced to make the solution dimensionally correct. Two special modes of the Neufville solution are well known. The mode $n = 0$ is the Oseen-Lamb vortex, which represents the viscous decay process of a singular line vortex. Second, the mode $n = 1$ leads to the Taylor vortex.

There is a transformation between the two families of the solutions:

$$\omega_n = \frac{A_0}{4\pi v} \left(\frac{2}{a}\right)^{\frac{1}{4}} (n+1) d_0^{-(n+1)} \bar{\omega}_{n+1}. \quad (52)$$

$$\Gamma_n(r,t) = \frac{A_0}{4\pi v} \left(\frac{2}{a}\right)^{\frac{1}{4}} d_0^{-(n+1)} \bar{\Gamma}_{n+1}. \quad (53)$$

A factor must be introduced to show the difference between these solutions as

$$G(n, v, A_0, d_0, a) = \frac{A_0}{4\pi v} \left(\frac{2}{a}\right)^{\frac{1}{4}} d_0^{-(n+1)}. \quad (54)$$

By setting a gauge condition

$$G(n, v, A_0, d_0, a) = 1. \quad (55)$$

we have

$$\Gamma_n(r,t) = \bar{\Gamma}_{n+1}(r,t). \quad (56)$$

It is important to note that under this gauge we still have

$$\omega_n = (n+1) \bar{\omega}_{n+1}. \quad (57)$$

This demonstrates that the new solution we found is completely different from the Neufville's solution.

# 6. Conclusions

In this paper, based on the generalized self-similar solutions, introducing an eigenvalue system, we made some further study on two-dimensional incompressible axisymmetric vortex flow. A new set of analytical solutions are obtained, and the Neufville (1957) solution is compared. At least there are different in the following aspects:

[1]The basic equation we are used in this paper is the evolution equation based on the circulation.

[2]Vortex field based on the new solution to repeat feature Taylor vorticity for the mode $n = 0$, rather than Neufville (1957) corresponding to Oseen vortex solutions.

[3] Even if with the gauge conditions we introduced, two kinds of solution, the vorticity field still exists huge difference.

[4]Eigenvalue system analysis method we propose can be easily extended to other vortex flow system.

It is note that the deeper causes of these difference call for further research.

**Acknowledgements:** The author would like to thank Prof. Wu Jie-Zhi and Zhang Shu-Hai for their useful discussion.

# Appendix 1：Normal equation

Consider a second-order homogenous linear equation in the general form:
$$f_2(x)y_{xx} + f_1(x)y_x + f_0(x)y = 0. \tag{a.1.1}$$

The substitution
$$y = u(x)exp\left\{-\frac{1}{2}\int \left(\frac{f_1}{f_2}\right)dx\right\}. \tag{a.1.2}$$

brings the equation (a.1.1) to the canonical ( or normal ) form
$$u_{xx} + f(x)u = 0. \tag{a.1.3}$$

Where
$$f(x) = \frac{f_0}{f_2} - \frac{1}{4}\left[\frac{f_1}{f_2}\right]^2 - \frac{1}{2}\left[\frac{f_1}{f_2}\right]_x. \tag{a.1.4}$$

# Appendix 2：Solution of the eigenvalue problem

Differential equation connected with the generalized Laguerre function is:
$$u_{xx} + \left(4n + 2\alpha + 2 - x^2 + \frac{1-4\alpha^2}{4x^2}\right)u = 0. \tag{a.2.1}$$

The corresponding solution is
$$u(x) = x^{\alpha+\frac{1}{2}}exp\left(-\frac{1}{2}x^2\right)L_n^{(\alpha)}(x^2). \tag{a.2.2}$$

wher  n = 0,1,2, … …